\documentclass[preprint]{aastex}

\begin{document}

\title{Stable Magnetic Fields in Static Stars}

\author{A. Gruzinov  }
\affil{CCPP, Physics Department, New York University, 4 Washington Place, New York, NY 10003}

\begin{abstract}
We prove that static fluid stars can stably support magnetic fields (within the ideal MHD approximation).
\end{abstract}

\section{Introduction}

Convectively unstable stars generate magnetic fields by a dynamo mechanism. When the fluid motion stops, should the field die out? 

A necessary condition for the existence of the magnetic field in a static star is the ideal magnetohydrodynamic (MHD) stability of the field. One might think that  ideal MHD stability of a given equilibrium magnetic field in a static star is an easy problem. The equation for small perturbations near the equilibrium is self-adjoint, and a variational principle that decides the stability question is easily formulated. But in practice (\citet{tay} and references therein) the variational principle has only been used  to prove that certain magnetic configurations are unstable.  Numerical simulations do suggest that mixed toroidal-poloidal fields can be stable \citep{bra}. 

Here we prove stability of a certain configuration --   mixed toroidal-poloidal field in a stably stratified incompressible ideally conducting star. The particular configuration that we prove stable is artificial -- our choice of the configuration was obviously driven by simplicity. However, "by continuity", a large class of stable fields close to our configuration should exist. In fact, the numerically stable fields of 
\citep{bra} are close to our configuration.

\section{Stable Magnetized Star}

\subsection{The initial field}

We will show that a stable equilibrium configuration exists which is close (in the sense specified below) to the  non-equilibrium {\it initial} configuration with the density field $\rho$ and magnetic field ${\bf B}$ given by
 \begin{equation}\label{dens}
\rho ( {\bf r} ) = \left\{
\begin{array}{cc}
\rho _1, & r<R\\
\rho _2, & R<r<R_s\\
0,          &  R_s<r
\end{array}\right.
\end{equation} 
 \begin{equation}\label{mfield}
{\bf B} ( {\bf r} ) = \left\{
\begin{array}{cc}
B_0\left( ~{2j_1(kr)\over kr} \cos \theta , ~ -(j_1'(kr)+{j_1(kr)\over kr})\sin \theta ,  ~j_1(kr)\sin \theta \right) , & r<R\\
0, & r>R
\end{array}\right.
\end{equation}
We use spherical coordinatization $(r,\theta , \phi )$;  $j_1$ and $j_1'$ are  the spherical Bessel function and its derivative;  $j_1(kR)=0$.

The magnetic field (\ref{mfield}) is the field of the minimal energy for a given helicity with a zero normal component on the surface of the sphere of radius $R$,
\begin{equation}\label{rot}
\nabla \times {\bf B}=k{\bf B}.
\end{equation} 
The importance of helicity for the problem of stability is well known and explained in the next subsection  (see \citet{arn} for discussion and references). 

The free parameters of the initial configuration are the radius of the star $R_s$, the radius of the magnetized region $R$, the densities of the magnetized and unmagnetized regions $\rho _1$ and $\rho _2$, and the amplitude of the field $B_0$. To simplify the proof of stability we assume that the stratification is strongly stable, $\rho_1\gg \rho_2$, and the magnetic field is weak,  $G\rho _1^2R^2\gg B_0^2$.

\subsection{The stable  field}

Assume incompressibility and ideal conductivity. Then an equilibrium configuration is an extremum of the total (gravitational plus magnetic) energy under incompressible deformations of the star and the magnetic field (comoving density constant, magnetic field frozen into the fluid). Stable equilibrium is the minimum of the energy under incompressible deformations.

The initial configuration is not an equilibrium, because the gravitational energy is at a minimum, while the magnetic energy is not extremal.  The Ampere force ${\bf j} \times {\bf B}$ does vanish inside the magnetized region due to (\ref{rot}). But there is a singular Ampere force on the boundary of the magnetized region at $r=R$. This force acts in the direction of deforming the sphere into an oblate spheroid. 
 
Consider the set of all configurations which can be obtained from (\ref{dens},\ref{mfield}) by incompressible deformations. We will show that the minimum-energy configuration belonging to the set is close to the initial configuration. In particular, the boundary of the magnetized region of the minimum-energy configuration is an axisymmetric oblate spheroid close to a sphere (in the mean square deviation sense).
 
The idea of the proof is straightforward. A deformation of the star causes a deformation of the boundary of the magnetized high-density region. Let $\psi (\theta , \phi )$ be the deviation of the deformed boundary from the sphere:
\begin{equation}\label{boundary}
r=(1+\psi (\theta , \phi ))R.
\end{equation} 
For $\rho_1\gg \rho_2$, the change of the gravitational energy $\delta W_g$ is a functional of $\psi$ only, after normalizing to the gravitational energy of  the initial density (\ref{dens}):
\begin{equation}
\delta W_g=|W_g| w_g[\psi ],
\end{equation} 
The change of the magnetic energy $\delta W_m$ depends on the shape of the boundary $\psi$ and on the deformation field inside the magnetized region. But since the frozen-in deformations conserve the magnetic helicity,  a lower bound on the change of the magnetic energy can be obtained by minimizing the magnetic energy for a given boundary (\ref{boundary}) and helicity. The boundary condition for this minimization is still the vanishing of the normal component of the magnetic field, because this property is preserved by the frozen-in deformation of the field. Thus we have 
\begin{equation}
\delta W_m>W_m w_m[\psi ],
\end{equation} 
where $W_m$ is the magnetic energy of the initial field (\ref{mfield}), and $w_m[\psi ]$ is the dimensionless change of the minimal magnetic energy for a given helicity due to the variation of the boundary $\psi$. 

The dimensionless  change of the gravitational energy $w_g[\psi]$ is positive, and as we show in Appendix I, up to the second order in $\psi$,
\begin{equation}
w_g[\psi ]>C_1I, ~~~~I\equiv \int d\Omega ~\psi ^2.
\end{equation} 
Here and below $C$ denotes positive dimensionless constants. It is assumed that the deformation $\psi$ does not contain a dipole mode (translation), for this mode clearly does not contribute to the magnetic energy, and will be stabilized by the non-zero outer density $\rho _2$.

The dimensionless change of the magnetic energy $w_m[\psi ]$ can be negative, but as we show in Appendix II, also up to the second order in $\psi$,
\begin{equation}
w_m[\psi ]>-(C_2I^{1/2}+C_3I)
\end{equation} 

The total energy of the minimum-energy configuration is smaller than the initial energy, 
 \begin{equation}
 \delta W_g+\delta W_m<0.
\end{equation} 
It follows that the minimum-energy configuration is close to the sphere:
\begin{equation}
 \int d\Omega ~\psi ^2~~<~~C{W_m^2\over W_g^2}~~\ll 1.
\end{equation} 
As shown in Appendix II,  the minimum-energy configuration is axisymmetric. The magnetic field of the minimum-energy configuration is close to the initial field (\ref{mfield}).

\section{Conclusion}
Thus static stars can stably support magnetic fields. Our field is hidden inside  the star, but it seems likely that a poloidal component which sticks through the surface of the star can be added without causing an instability, at least a component with a small enough amplitude. This is because the minimum-energy state that we have presented forbids any deformations of the magnetized region other than motions along the magnetic surfaces. Therefore, adding a weak net poloidal flux which goes through the surface of the star and through the center of the star is likely to cause just a small deformation of our equilibrium -- one cannot kill this poloidal component by the motions along the magnetic surfaces.

\acknowledgments

I thank Peter Goldreich, Jeremy Goodman, Andrew MacFadyen and Boris Khesin for discussions. This work was supported by the David and Lucile Packard foundation.

\appendix

\section{Gravitational Energy}
Take $R=1$, $\rho _1=6$, $\rho _2=0$. Then the gravitational potential is
 \begin{equation}
\phi =  \left\{
\begin{array}{cc}
-3+r^2+\sum a_{lm}r^lY_{lm} , & r<1+\psi \\
-2r^{-1}+\sum b_{lm}r^{-(l+1)}Y_{lm} , & r>1+\psi ,
\end{array}\right.
\end{equation}
where $a$, $b$ are constants, $Y$ are the spherical functions.  From continuity of the potential and its derivative across the boundary $r=1+\psi$, we get, to first order in $\psi$,
\begin{equation}\label{lin}
a_{lm}=b_{lm}={-6\over 2l+1}\psi _{lm} ,
\end{equation}
where 
\begin{equation}
\psi=\sum \psi _{lm} Y_{lm}.
\end{equation}
Note that $\psi _0$ is a second order quantity -- from volume conservation, to second order in $\psi$,
\begin{equation}
\psi _0={1\over 2\sqrt{\pi } }\sum_{l\geq 1} |\psi _{lm}|^2.
\end{equation}
To second order in $\psi$, the continuity of the potential gives
\begin{equation}
a_0Y_0+\sum_{l\geq 1} (a _{lm}-b_{lm}) Y_{lm}=3\psi ^2.
\end{equation}
which gives 
\begin{equation}\label{a0}
a_0={3\over 2\sqrt{\pi} }\sum |\psi _{lm}|^2.
\end{equation}

The gravitational energy is 
\begin{equation}
W_g={1\over 2}\int d^3r~\rho \phi =3\int_0^{1+\psi}r^2~dr~\sin \theta ~d\theta ~d\phi (-3+r^2+\sum a_{lm}r^lY_{lm}).
\end{equation}
The change of energy is, up to second order in $\psi$,
\begin{equation}
w_g[\psi ] \propto \int \sin \theta ~d\theta ~d\phi \left( \psi +2\psi ^2+\sum a_{lm}Y_{lm}({1\over l+3}+\psi )\right) =\sum (|\psi _{lm}|^2+\psi _{lm}a_{lm}^*)+{2\sqrt{\pi }\over 3}a_0,
\end{equation}
and using (\ref{lin}, \ref{a0}), we get the final answer
\begin{equation}\label{wg}
w_g[\psi ] \propto \sum_{l\geq 1} (1-{3\over 2l+1})|\psi _{lm}|^2.
\end{equation}
In the $\rho _2=0$ approximation, there is no penalty for the simple displacement of the high-density region (the dipole, $l=1$,  modes). A non-zero outer density adds a positive contribution from the dipole.  However, we don't need to include this contribution, because the dipole does not contribute to the magnetic energy $w_m[\psi ]$.

\section{Magnetic Energy}
Consider magnetic fields that minimize magnetic energy, $W\equiv \int d^3r B^2$, for a given helicity, $H\equiv \int d^3r{\bf A}\cdot {\bf B}$, where $\nabla \times {\bf A}={\bf B}$. The minimizing fields are solutions of $\nabla \times {\bf B}=k{\bf B}$. The minimizing field of the sphere is (\ref{mfield}). When one deforms the sphere without changing the enclosed volume, the value of the Lagrange multiplier $k$ changes, and from $W=kH$, one gets the new minimal energy for a given helicity. 

In our calculation of the minimal energy we prefer to keep $k$ constant by allowing  volume changing deformations of the boundary. Then, by a uniform expansion or contraction, we restore the volume, and calculate the minimum energy from the scaling 
\begin{equation}\label{scale}
W=CHV^{-1/3},
\end{equation}
where $C$ depends on the shape of the boundary, but not on the enclosed  volume $V$. 

Thus, we will calculate magnetic field ${\bf B}$ satisfying 
\begin{equation}\label{beq}
\nabla \times {\bf B}={\bf B}, ~~r<x_1+\psi (\theta ,\phi ),
\end{equation}
and tangent to the boundary $r=x_1+\psi (\theta ,\phi )$. 
Here $x_1$ is the first zero of $j_1$, and the unperturbed, zeroth order in $\psi$,  solution is 
\begin{equation}\label{b0}
{\bf B}_0=\left( ~{2j_1\over r} \cos \theta , ~ -(j_1'+{j_1\over r})\sin \theta ,  ~j_1\sin \theta \right),
\end{equation}
The exact solution for the arbitrary change of the boundary $\psi$ (with an appropriate uniform expansion harmonic $\psi _0$ to keep $k=1$) is given by (\ref{beq}):
\begin{equation}\label{b}
{\bf B}={\bf B}_0+\sum B_{lm}\left( ~l(l+1){j_l\over r} Y_{lm} , ~ {im\over \sin \theta }j_lY_{lm}+(j_l'+{j_l\over r})\partial _\theta Y_{lm} ,  ~-j_l\partial _\theta Y_{lm}+{im\over \sin \theta }(j_l'+{j_l\over r})Y_{lm}\right).
\end{equation}

The field must be tangent to the boundary:
\begin{equation}
B_r={1\over r}B_\theta \partial _\theta \psi+{1\over r\sin \theta }B_\phi \partial _\phi \psi, ~~r=x_1+\psi (\theta ,\phi ),
\end{equation}
or, up to second order in $\psi$,
\begin{equation}\label{basic} 
\begin{array}{c}
\sum B_{lm}l(l+1)j_lY_{lm}=  -j_1'(1-{\psi \over x_1}){1\over \sin \theta }\partial _\theta (\sin ^2\theta \psi )+j_1'\psi \partial _\phi \psi   \\
- \psi \sum B_{lm}l(l+1)j_l'Y_{lm}+\partial _\theta \psi\sum B_{lm}\left( j_l{im\over \sin \theta}Y_{lm}+(j_l'+{j_l\over x_1})\partial _\theta Y_{lm}\right) \\
+ {1\over \sin \theta }\partial _\phi \psi \sum B_{lm}\left(- j_l\partial _\theta Y_{lm}+(j_l'+{j_l\over x_1}){im\over \sin \theta}Y_{lm}\right).
\end{array}
\end{equation}
The argument of Bessel functions here and below is $x_1$. To first order (\ref{basic}) gives 
\begin{equation}\label{basic1} 
\sum B_{lm}l(l+1)j_lY_{lm}=-j_1'{1\over \sin \theta }\partial _\theta (\sin ^2\theta \psi ),
\end{equation}
which allows to express the $B$ harmonics in terms of $\psi$ harmonics to first order. Using the formula for the derivative of spherical functions
\begin{equation}
\sin \theta \partial _\theta Y_{lm}=l\left({(l+1)^2-m^2\over 4(l+1)^2-1}\right) ^{1/2}Y_{l+1,m}- (l+1)\left({l^2-m^2\over 4l^2-1}\right) ^{1/2}Y_{l-1,m},
\end{equation}
and integrating (\ref{basic1}) multiplied by $Y_{lm}^*$ over angles we get
\begin{equation}
B_{lm}={j_1'\over j_l}\chi _{lm},~~~~
\chi _{lm}\equiv {1\over l+1}\left({(l+1)^2-m^2\over 4(l+1)^2-1}\right) ^{1/2}\psi _{l+1,m}- {1\over l}\left({l^2-m^2\over 4l^2-1}\right) ^{1/2}\psi _{l-1,m}.
\end{equation}
The first-order result (\ref{basic1}) allows to re-write (\ref{basic}), to second order, in the following useful form
\begin{equation}\label{main}
\sum B_{lm}l(l+1)j_lY_{lm}={1\over \sin \theta }\partial _\theta F_1+{1\over \sin ^2 \theta }\partial _\phi F_2,
\end{equation}
where 
\begin{equation}
F_1=(-j_1'\sin ^2\theta +\sin \theta \partial _\theta G_2+\partial _\phi G_1)\psi,~~~
F_2=({1\over 2}j_1'\sin ^2\theta \psi -\sin \theta \partial _\theta G_1+\partial _\phi G_2)\psi
\end{equation}
\begin{equation}
G_1=\sum B_{lm}j_lY_{lm}, ~~~G_2=\sum B_{lm}(j_l'+{j_l\over x_1})Y_{lm}.
\end{equation}

Choosing different values for $B_{lm}$, one gets functions with arbitrary $l>1$ harmonics but with vanishing $l=0$ and $l=1$ harmonics in the left-hand side of (\ref{main}). The right-hand side of (\ref{main}) automatically gives zero for the $l=0$ harmonic. The vanishing of the $l=1$, $m=0$ harmonic means that the $l=0$ harmonic of $F_1$ vanishes. This gives, to second order in $\psi$, 
\begin{equation}\label{psi}
\psi _0 = {1\over \sqrt{5} }\psi _{2,0}+{3\over 4\sqrt{\pi }} \sum \left( ({1\over x_1}+{j_l'\over j_l})l(l+1)|\chi _{lm}|^2+im\psi_{lm}^*\chi _{lm}\right).
\end{equation}
We do not need the $l=1$, $m=\pm 1$ equation, for it simply constrains the orientation of the unperturbed field.

Equation (\ref{psi}) describes the deformations of the sphere, which leave the minimal energy for a given helicity unchanged: $W=H$, because $\nabla \times {\bf B}={\bf B}$. However the deformation (\ref{psi}) changes the volume inside the boundary: the volume inside   $r=x_1+\psi (\theta ,\phi )$ is greater than the volume inside the sphere  $r=x_1$ by 
\begin{equation}\label{vol}
\delta V \propto 2\sqrt{\pi}x_1\psi _0 +\sum |\psi _{lm}|^2.
\end{equation}

Restoring the volume by a uniform contraction, we get the change of the minimal energy per unit helicity caused by the $\psi$-deformation. From (\ref{scale}): $w_m[\psi ]\propto \delta V$.  Using (\ref{psi}) in (\ref{vol}), we obtain the final expression
\begin{equation}\label{fin}
w_m[\psi ]\propto {2\sqrt{\pi }\over \sqrt{5} }x_1\psi _{2,0}+\sum \left( |\psi _{lm}|^2+{3\over 2}x_1im\psi _{lm}^*\chi _{lm}+{3\over 2}(1+{x_1j_l'\over j_l})l(l+1)|\chi _{lm}|^2\right).
\end{equation}

The linear term of this functional, $\propto \psi _{2,0}$, says that magnetic energy can be reduced by a deformation with a negative $(2,0)$ harmonic, that is by an oblate  deformation $\propto 1-3\cos ^2\theta$. 

For $l\gg 1$, the quadratic term in (\ref{fin}) is non-negative (minimize over $\chi$ as if $\chi$ were independent, using $j_l(x)\propto x^l$). Therefore, one has a lower bound: $w_m[\psi ]>-(C_2I^{1/2}+C_3I)$, $I\equiv \int d\Omega ~\psi ^2$.

Modes with different $m$ are  not coupled to each other in (\ref{fin}); it follows that the minimal energy state is axisymmetric.

\clearpage

\end{document}